\documentclass[letterpaper]{article}

\usepackage{natbib,alifeconf}

\usepackage{amsmath}
\usepackage{amssymb}
\usepackage{xcolor}

\newcommand\Tstrut{\rule{0pt}{2.6ex}}         

\newcommand\asspos{\mathrel{\stackrel{\makebox[0pt]{\mbox{\normalfont\tiny +}}}{\Leftrightarrow}}}

\newcommand\assposforward{\mathrel{\stackrel{\makebox[0pt]{\mbox{\normalfont\tiny +}}}{\Rightarrow}}}

\newcommand\assneg{\mathrel{\stackrel{\makebox[0pt]{\mbox{\normalfont\normalsize -}}}{\Leftrightarrow}}}

\newcommand\assnegforward{\mathrel{\stackrel{\makebox[0pt]{\mbox{\normalfont\normalsize -}}}{\Rightarrow}}}

\title{Agent Heterogeneity Mediates Extremism in an Adaptive Social Network Model}

\author{Seth Bullock$^{1}$ and Hiroki Sayama$^{2,3}$\\
\mbox{}\\
$^1$ Computer Science, University of Bristol, UK\\
seth.bullock@bristol.ac.uk\\
$^2$ Center for Collective Dynamics of Complex Systems, Binghamton University, Binghamton, NY 13902-6000, USA\\
$^3$ Waseda Innovation Lab, Waseda University, Shinjuku, Tokyo 169-8050, Japan}

\begin{document}

\maketitle

\begin{abstract}

An existing model of opinion dynamics on an adaptive social network is extended to introduce \emph{update policy heterogeneity}, representing the fact that individual differences between social animals can affect their tendency to form, and be influenced by, their social bonds with other animals. As in the original model, the opinions and social connections of a population of model agents change due to three social processes: conformity, homophily and neophily. Here, however, we explore the case in which each node's susceptibility to these three processes is parameterised by node-specific values drawn independently at random from some distribution. This introduction of heterogeneity increases both the degree of extremism and connectedness in the final population (relative to comparable homogeneous networks) and leads to significant assortativity with respect to node update policy parameters as well as node opinions. Each node's update policy parameters also predict properties of the community that they will belong to in the final network configuration. These results suggest that update policy heterogeneity in social populations may have a significant impact on the formation of extremist communities in real-world populations.





\end{abstract}

\section{Introduction}

When living creatures form social groups, the composition of these communities tends to reflect the shared traits of the social agents involved. For example, an agent's relatively stable physical or socio-economic properties such as its spatial location or its economic class or occupation may influence the social groups that it forms or joins. Simultaneously, the groups to which an agent belongs may tend to shape some of these key traits by structuring the kinds of influence to which it is subjected. An agent might change its location in order to remain part of a herd, for instance, or spend its money in a way that is influenced by the members of its socio-economic group. 

Beyond the influence of these relatively stable traits, community formation may also be driven by more volatile agent characteristics such as the extent to which individual agents share opinions or attitudes. For instance, researchers have sought to detail the role of social processes in the formation of political ideologies \citep{prior2013media,morales2015measuring},  religious extremism \citep{manrique2018generalized,badawy2018rise}, healthcare choices \citep{kata2012anti}, dietary preferences \citep{cole2011vegaphobia,reilly2016gluten} and issues related to technologies and innovation \citep{coccia2016radical,naranjo2017organizational}. Analogously, engineers interested in designing or managing distributed multi-agent systems may have similar interests in understanding flows of influence within swarms of collaborating robots \citep{swarmintel2016,alifexv-lenka,swarmintel2018} or populations of interacting software agents \citep{eps267064,eps342217}.


Across these different settings, community formation can be understood as an ongoing reflexive process of coevolutionary adaptation both \emph{on} and \emph{of} a social network, i.e., the network constrains which agents are more likely to interact with one another and, simultaneously, is shaped and reshaped by these interactions \citep{eps271552,sayama2020enhanced}. To gain insight into such systems, researchers have studied theoretical models of adaptive social networks, where network topology and node traits co-evolve simultaneously \citep{eps266036,gross2009adaptive,sayama2013modeling}. Several papers have focused on phase transitions between connected and fragmented network topologies \citep{holme2006nonequilibrium,zanette2006opinion,kozma2008consensus,bohme2011analytical,sayama2020beyond}, while others have studied global drift phenomena in social diffusion \citep{sayama2015social}. 

Within such models, the manner in which agents alter their opinions and their connections with other agents is typically governed by a single update policy that is adopted by all agents, i.e., agents are behaviourally \emph{homogeneous}. For instance, \cite{sayama2020alife} introduced a model of opinion dynamics on an adaptive social network in which all agents were subject to some degree of social conformity (adjusting opinions in the direction of the local social norm), homophily (strengthening connections to agents with similar opinions) and neophily (strengthening connections to agents whose opinions are novel with respect to the local social norm). The strength of each of these factors was varied systematically over a series of numerical simulations in order to evaluate their impact on network dynamics, but for each numerical simulation, the strength of each of these three factors was homogeneous across all agents in the population. 

The study found that to the extent that the population was strongly homophilic, it tended to fragment into a relatively large number of communities with agents from the same community tending to share a similar opinion, but agents from different communities exhibiting divergent opinions that could be extreme relative to the population average. By contrast, a strongly neophilic population tended to  form a small number of large communities (often just one), featuring opinions that were moderate and fairly homogeneous (but diverse by comparison with the tightly converged opinions within a single typical homophilic community). Subsequent work demonstrated that high social conformity was also capable of preventing social fragmentation even under conditions of low neophily \citep{sayama2022ppam}.

It is known that introducing individual-level behavioral diversity within multi-agent models can generate nontrivial macroscopic outcomes \citep{swarmchem,sayama2020beyond,bennett-alife2022} and that ``personality differences'' can have adaptive significance for biological populations \citep{dall2004}. However, to date, the influence of agent heterogeneity on opinion dynamics has tended only to be explored in the context of simple bounded confidence models \cite[e.g.,][]{hegselmann&krause2002}. Here, two agents must hold opinions that differ by less than some threshold amount in order for them to interact with each other. Studies have explored the impact of allowing this threshold to vary between agents on the time for opinions to converge or to polarise \citep[e.g.,][]{lorenz2007, liang2013, cheng2019, kan2023}.

Sayama's (\citeyear{sayama2020alife}) model differs from these bounded confidence models in considering a homogeneous population of agents that are subject to a more complex set of parameterised social processes. This raises the following question: how sensitive is this model to the assumption that all agents obey the same update policy, responding to the same social pressures in the same way and to the same extent? 

Here we extend Sayama's (\citeyear{sayama2020alife}) homogeneous agent-based model of opinion dynamics on an adaptive social network to explore scenarios in which agents are \emph{heterogeneous} in the parameters of the node-specific update policies governing the adaptation of their opinions and their connections with other agents. We conduct a series of numerical simulations and perform regression analyses to elucidate the effects of this update policy heterogeneity on a population's tendency towards extremist individuals and communities.

\section{Model}

Following \cite{sayama2020alife}, the opinions and social structure of a population of agents is represented by a fully-connected graph, $G$, comprising a set of nodes, $V$, connected by a set of weighted, directed edges, $E$, (excluding self-connections), where $|V|=N$ and $|E| = N(N-1)$. Each node $i \in V$ represents an individual agent with an opinion $x_i \in \mathbb{R}$. Each node is influenced by each of its network neighbours. The strength of the influence exerted by node $j$ on node $i$ is represented by the weight of network edge $w_{ij} \in \mathbb{R}_{\geq 0}$.

Over time, each node's opinion has a tendency to shift towards the weighted average opinion of its local social neighborhood (social conformity) and also to drift at random (noise). Each node's incoming edge weights also change over time such that they tend to reflect the extent to which the node and its upstream neighbour share a similar opinion (homophily), and also tend to reflect the extent to which the upstream neighbour's opinion is distinct from the weighted average opinion of the node's social community (neophily). That is, opinions and edge weights co-evolve over time through four mechanisms: (1) social conformity, (2) noise, (3) homophily and (4) neophily. 

These dynamics are determined as follows:
\begin{align}
\frac{d x_i}{d t} &= c_i \left( \langle x \rangle_i - x_i \right) + \epsilon \\
\frac{d w_{ij}}{d t} &= h_i F_h(x_i, x_j) + a_i F_a(\langle x \rangle_i, x_j) \\
\langle x \rangle_i &= \frac{\sum_{j \in N_i} w_{ij} x_j}{\sum_{j \in N_i} w_{ij}}
\end{align}
Here, $N_i$ is the set of in-neighbors of node $i$ (i.e., all nodes apart from $i$ itself); $\langle x \rangle_i$ is the local weighted average opinion, or social norm, perceived by node $i$; $\epsilon$ represents a stochastic fluctuation term that influences node opinions; and $c_i$, $h_i$, and $a_i$ are node-specific parameters that determine, respectively, the strength of social conformity, homophily, and neophily specific to node $i$ (which is the model's novel element). Behavioural functions $F_h$ and $F_a$ determine the rate of edge weight change based on opinion distance, defined as follows:
\begin{align}
F_h(x_i, x_j) &= \theta_h - |x_i - x_j| \\
F_a(\langle x \rangle_i, x_j) &= |\langle x \rangle_i - x_j| - \theta_a
\end{align}
Here $\theta_h$ and $\theta_a$ are fixed population-wide parameters that act as threshold opinion distances separating a regime in which weights from an upstream neighbour are strengthened from a regime in which they are weakened.

Note that $w_{ij}$ is bounded to be non-negative, i.e., any negative values are rounded up to zero. Where a node's incoming edge weights are all zero, its local weighted average opinion is undefined and neither the influence of conformity nor neophily are applicable.

In previous work \citep{sayama2020alife}, the influence of \emph{homogeneous} social conformity, homophily and neophily on network formation was explored by varying the associated model parameters \emph{between} independent simulation runs. During any single instantiation of the model, nodes were homogeneous in their update parameters, i.e., $\forall i \in V : c_i=c, h_i=h, a_i=a$. By contrast, for the results presented here, node-specific values for these three node parameters were each drawn from a parameter-specific random distribution during initialisation and were kept fixed for the duration of each network simulation. In combination, the $c_i$, $h_i$ and $a_i$ values of a specific node are referred to as its \emph{update policy} and the population mean parameter values, $\mu_c$, $\mu_h$, and $\mu_a$ define the network's \emph{mean update policy}. 

We implemented the above adaptive social network model in Python 3.8 with the NetworkX package.\footnote{The simulator code is available from the corresponding author upon request.}

\section{Simulations}

\subsection{Simulation Settings}


For all simulation runs reported here: $N = 1000$, $\theta_h = 0.03$, and $\theta_a = 0.03$. The initial configuration of the network was such that every ordered pair of unique nodes was connected by a directed edge with a weight randomly sampled from the uniform distribution $[0,1]$ and each node had a random opinion sampled from the standard normal distribution ${\cal N}(0, 1^2)$. 

Node-specific parameter values were assigned using one of two methods. Simulations exploring \emph{Uniform Heterogeneity} were initialised such that for each node, $i$, values of $c_i$, $h_i$ and $a_i$ were randomly sampled from the uniform distribution $[0.01, 0.3]$, i.e., the original range of parameter values explored by \cite{sayama2020alife}. For simulations exploring \emph{Non-uniform Heterogeneity}, each $c_i$ value was drawn from ${\cal N}(\mu_c, \sigma_c^2)$, $h_i$ from ${\cal N}(\mu_h, \sigma_h^2)$ and $a_i$ from ${\cal N}(\mu_a, \sigma_a^2)$. Below, we report results for $\mu_c=\mu_h=0.05$, $\mu_a=0.25$ and $\sigma_c=\sigma_h=\sigma_a=0.025$. 

Each simulation instantiation employed a simple Euler forward integration method with time interval $\Delta t = 0.1$ for $t$ running from $0$ to $100$. The stochastic effect of $\epsilon$ was simulated by adding a random number sampled from ${\cal N}(0, 0.1^2)$ to each $x_i$ at every interval $\Delta t$.

\subsection{Community Structure and Assortativity}

The Louvain modularity maximization method \citep{blondel2008fast} was employed to assign each node in the final network configuration to one of a set of non-overlapping communities. This method requires that the network be undirected. Consequently, after each simulation run was completed, an undirected network $G'$ was constructed to be the equivalent of the final state of network $G$ at $t = 100$. The weight of each undirected edge in $G'$ was set to be equal to the mean weight of the directed edges between the same pair of nodes in $G$, i.e., $w_{ij}^{G'} = \frac{1}{2}(w_{ij}^G + w_{ji}^G)$. The community structure of $G'$ was then determined using the Louvain modularity maximization method and the resultant assignment of nodes to communities was applied to $G$ for the purposes of all subsequent analyses. 

While assortativity is often taken to refer to the tendency of a network's nodes to be connected to neighbours with similar degree, here we will consider the assortativity of several network properties, e.g., to what extent does the weight from $j$ to $i$ tend to vary systematically with the difference between their opinions, $x_j$ and $x_i$, or between their conformity parameters, $c_j$ and $c_i$, or indeed between the homophily $h_j$ of $j$ and the neophily $a_i$ of $i$. In each case, the assortativity of the final network was calculated using the method presented by \cite{yuan2021} which, unlike standard assortativity measures, is designed to deal with networks that are both weighted and directed.

\subsection{Outcome Measures}

Our first focus is on the overall network-level impact of introducing heterogeneity. We measure this in terms of the difference between the final configurations of networks formed under conditions of Uniform Heterogeneity and those formed under comparable Homogeneous conditions, i.e., where the Homogeneous agents all employ an update strategy that is the mean of those employed by the Heterogeneous agents, $\langle\mu_c, \mu_h, \mu_a\rangle$. 

Subsequently, we are interested to discover the influence of a node's update policy on the outcome of network dynamics. This outcome can be understood in terms of the node's influence on its own final \emph{individual} properties (e.g., its final opinion, its final strength, etc.), on the properties of the \emph{community} that the node finds itself within (e.g, the size of the community, the average opinion of the community), and on the overall properties of the entire \emph{network} (e.g., how assorted is the network with respect to $x_i$, $c_i$, $h_i$, or $a_i$). 

\subsubsection{Individual Node Measures}

Below we consider the relationship between a node's update policy and several node-level measures. A node's \emph{in-strength} and \emph{out-strength} represent, respectively, the sum of a node's incoming weights, and the sum of its outgoing weights. A node's \emph{community in-strength} and \emph{community out-strength} restricts the previous two measures to consider only neighbours within a node's own community. Finally, a node's \emph{deviance} is the absolute difference between its opinion and the network average opinion, whereas its \emph{within-community deviance} is the absolute difference between its opinion and the average opinion within its community.  

\subsubsection{Community Measures}

Below we consider the impact of a node's update policy on its community's \emph{size} and \emph{mean edge weight} and the \emph{range}, \emph{standard deviation} and \emph{deviance} of the opinions of the agents within it. A community's deviance is measured as the absolute difference between the average opinion of the agents within it and the network average opinion and can be considered a measure of the extent to which a community contains extreme members.

\begin{figure*}[t]
\minipage{0.32\textwidth}
  \includegraphics[width=\linewidth,trim={2.5cm 2.5cm 2.5cm 2.5cm},clip]{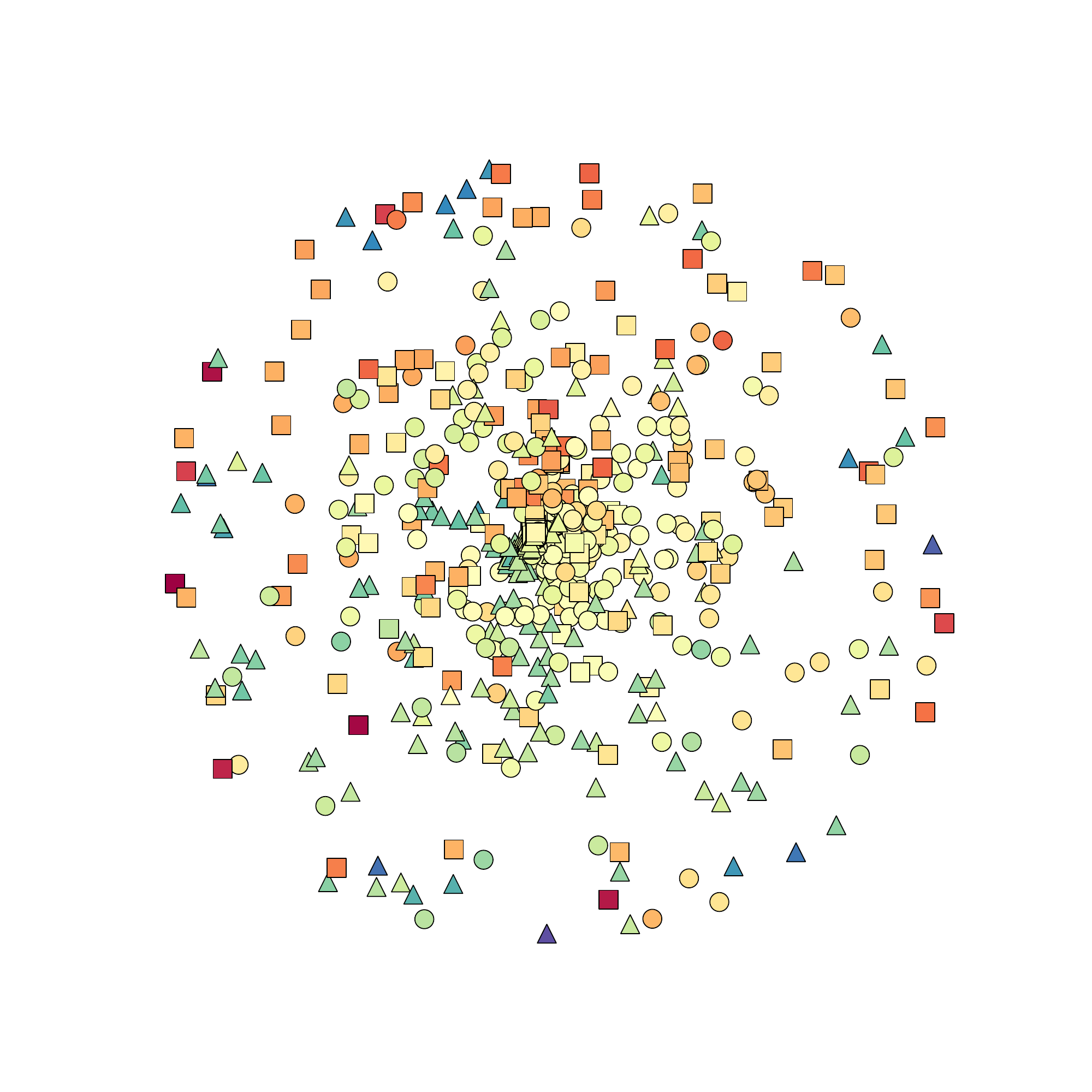}
\endminipage\hfill
\minipage{0.32\textwidth}%
  \includegraphics[width=\linewidth,trim={2.5cm 2.5cm 2.5cm 2.5cm},clip]{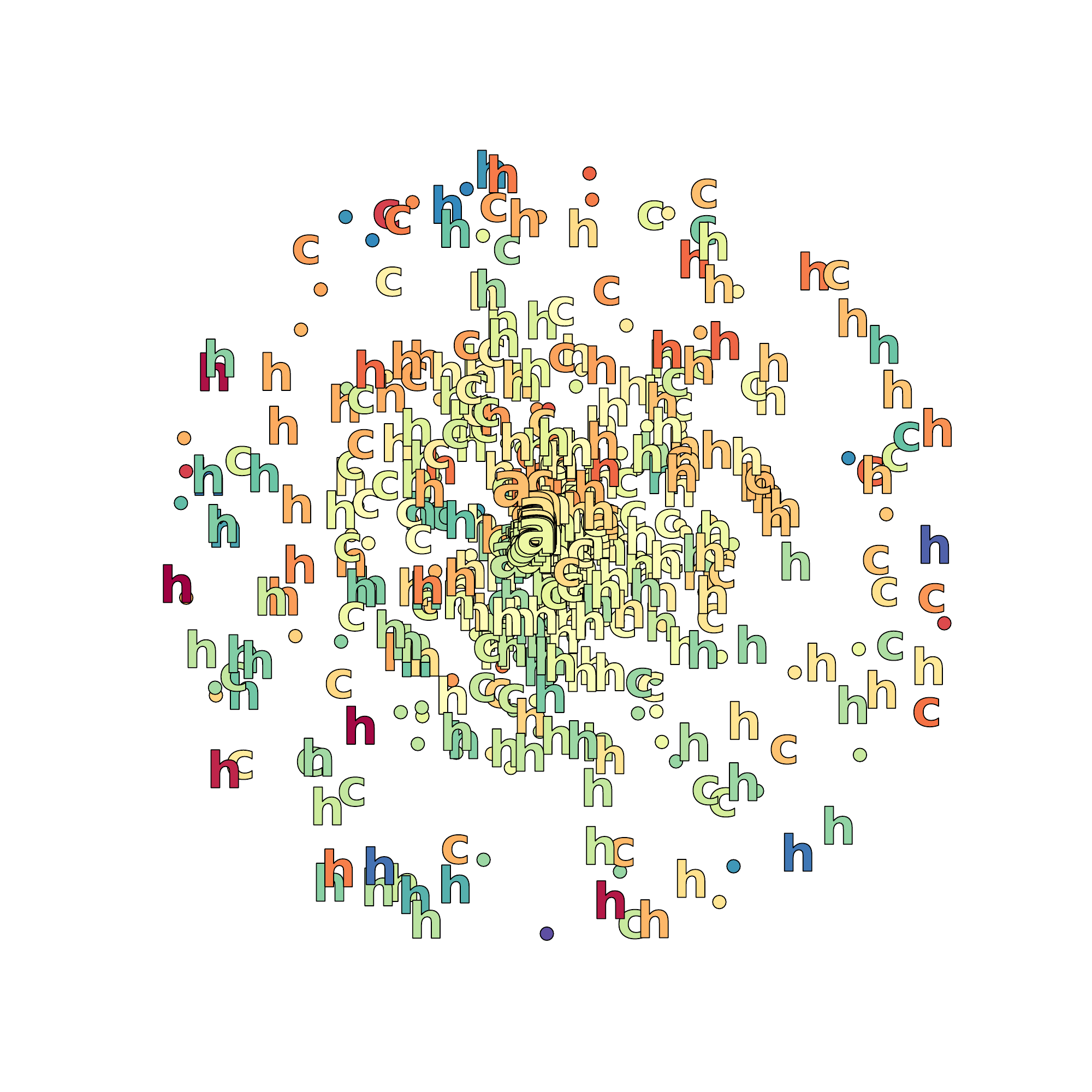}
\endminipage\hfill
\minipage{0.32\textwidth}
  \includegraphics[width=\linewidth,trim={2.5cm 2.5cm 2.5cm 2.5cm},clip]{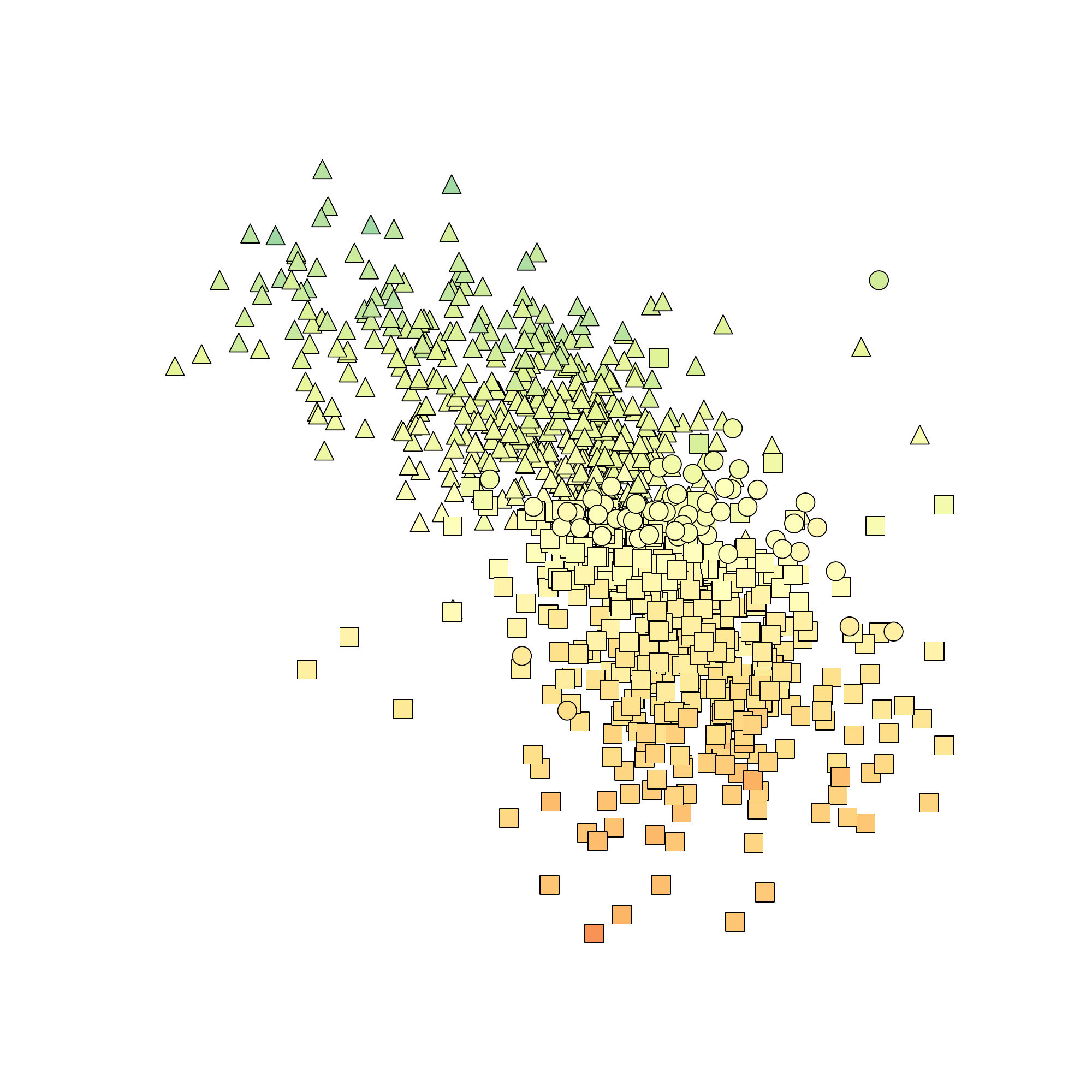}
\endminipage\hfill

\caption{Network visualizations depict \emph{(left)} a representative network formed under conditions of Uniform Heterogeneity with marker shape indicating community membership, \emph{(centre)} the same network with letters indicating the highest policy parameter for nodes with relatively extreme update policies (i.e., where any update parameter is greater than 0.2), and \emph{(right)} a comparable Homogeneous network. In all cases, nodes are coloured by their opinion (using a consistent colour scale) and node placement simulates the effect of (invisible) spring-like edge weights using a standard Fruchterman-Reingold force-directed algorithm.}
\label{fig:networks}

\end{figure*}

\begin{figure*}[t]
  \includegraphics[width=\linewidth,trim={0cm 0cm 0cm 0cm},clip]{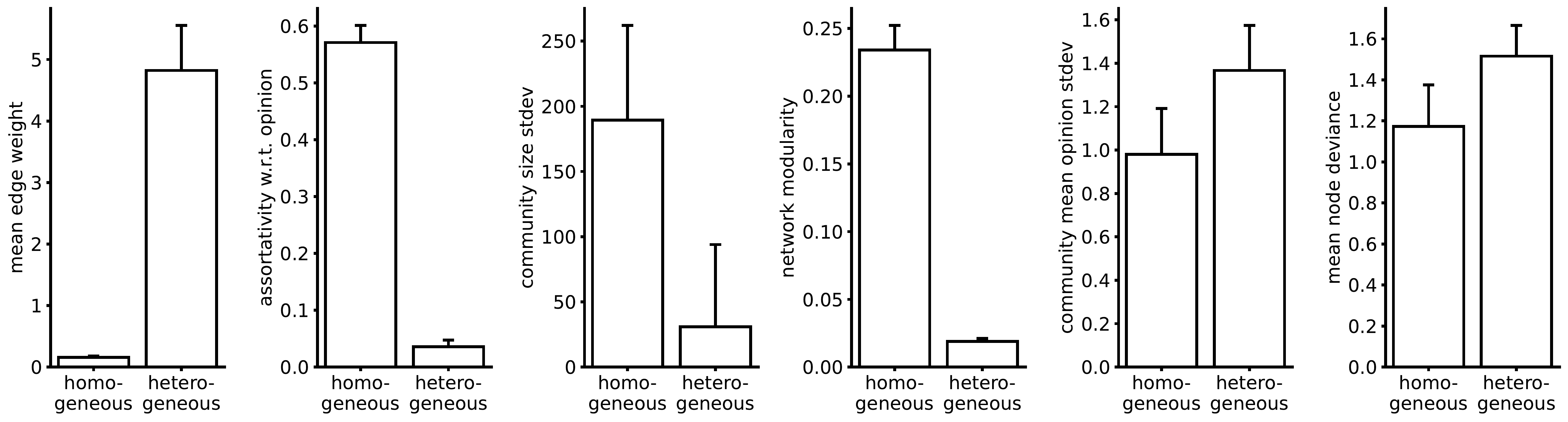}
\caption{Comparison between properties of 10 networks formed under conditions of Uniform Heterogeneity and 10 comparable Homogeneous networks. Error bars show standard deviations. Differences between means are all significant (two-tailed unequal variance t-tests, $p<10^{-7}$ in all cases).}
\label{fig:barcharts}

\end{figure*}

We are also able to consider the impact of a node's update policy on the tendency for its community to be \emph{assorted} in various ways. Positive/negative assortativity on \emph{opinion} measures the extent to which nodes within a community tend systematically to be strongly connected to neighbours with a similar/dissimilar opinion. Likewise, positive/negative assortativity on, e.g., \emph{conformity} measures the extent to which nodes within a community tend systematically to be strongly connected to neighbours with a similar/dissimilar conformity parameter, $c_i$.

\subsubsection{Population Measures}

We also report assortativity measures at the level of the entire population, e.g., are nodes assorted on their strength, or their $c_i$, $h_i$ and/or $a_i$ values?

\section{Results}

\subsection{Uniform Heterogeneity}

First, we assess the overall impact of introducing heterogeneity by comparing the properties of networks formed under conditions of Uniform Heterogeneity with the properties of comparable networks formed under Homogeneous conditions in which every node shares the same update policy (which is the mean of those employed by the Uniform Heterogeneous nodes). 

Figure~\ref{fig:networks}\emph{(left)} depicts a typical network formed under conditions of Uniform Heterogeneity. Nodes with similar opinions (colour) tend to be placed relatively close to each other reflecting some positive assortativity of nodes with respect to their opinion, and each of the three identified communities (represented by square, circle and triangle markers, respectively) is associated with a somewhat characteristic range of opinion (colour). However, these effects are not strong and there is evidence of considerable inter-community connectivity and variability, indicated by the wide spatial distribution of nodes and lack of correlation between a node's placement and its opinion or community. 

In contrast, the comparable Homogeneous network depicted in Figure~\ref{fig:networks}\emph{(right)} is more strongly assorted on opinion (indicated by the consistent colour gradient across the network), exhibits a stronger association between each community and a characteristic range of opinion, and shows less evidence of inter-community connectivity and variability. However, the absence of nodes with colours at the extreme ends of the colour map (dark red and dark blue) indicates that the range of opinion expressed across the Homogeneous network is not as great as in the Heterogeneous network and that there is greater similarity between the average opinions of the three communities. 

The central plot of Figure~\ref{fig:networks} shows that highly homophilic ({\bf h}) and highly conformist ({\bf c}) nodes can be found across the network and are not associated with particular opinions or communities. However, highly neophilic nodes (represented by {\bf a} symbols) are all tightly clustered in the very centre of the network indicating that these nodes are strongly connected to each other and that they may have a role in linking together disparate parts of the network.

Figure~\ref{fig:barcharts} confirms that Uniform Heterogeneous networks have stronger average edge weights than comparable Homogeneous networks and are less strongly assorted with respect to node opinion, and that while both sets of networks all exhibited the same number of communities (three), for Heterogeneous networks these communities are more regular in size, have weaker modularity, and have average opinions that are more varied, meaning that their node opinions deviate further from the network average. In summary, Uniform Heterogeneity leads to networks that are less assorted with respect to node opinion than comparable Homogeneous networks, and less fragmented with respect to network structure, but exhibit increased extremism at the level of the individual and the community. 

Note that this tendency to result in wider opinion diversity across the network's communities could be regarded as positive in the context of, e.g., the persistence of different cultures, languages or academic sub-disciplines in the face of pressure towards homogenization, but is often presented in terms of more negative issues such as political polarisation or ideological extremism in the opinion dynamics literature \citep[e.g.,][]{levin2021}.

Tables 1, 2 and 3 present the results of multiple linear regression models that capture the influence of node update policy parameters on the properties of 10 final network configurations formed under conditions of Uniform Heterogeneity, while Table 4 presents mean assortativity metrics for the same 10 networks. A brief summary of the main significant effects is provided below.

\begin{table*}[tbph]
\caption{Multiple linear regression of a node's \emph{individual-level} outcome measures on its update policy parameters ($c_i$, $h_i$, $a_i$) and their interactions (bottom three rows). Models were built on data from $10^4$ nodes obtained from $10\times$1000-node networks formed under conditions of \emph{Uniform Heterogeneity}. Statistically significant coefficients are indicated with asterisks (*: $p < 10^{-2}$; **: $p < 10^{-3}$; ***: $p < 10^{-4}$; etc.).}
\[
\begin{array}{c|l|l|l|l|l|l}
\hline
\text{Outcome} & \text{ } & \multicolumn{1}{c|}{\text{within-community}} & \text{ } & \multicolumn{1}{c|}{\text{within-community}} & \text{ } & \multicolumn{1}{c}{\text{within-community}} \\
\text{measure} & \multicolumn{1}{c|}{\text{in-strength }} & \multicolumn{1}{c|}{\text{in-strength}} & \multicolumn{1}{c|}{\text{out-strength}} & \multicolumn{1}{c|}{\text{out-strength}} & \multicolumn{1}{c|}{\text{deviance}} & \multicolumn{1}{c}{\text{deviance}}\\
\hline
\text{const.}	&	-5571^{\text{****}}	&	-1770^{\text{****}}	&	\phantom{-}5134^{\text{****}}	&	\phantom{-}1838^{\text{****}}	&	\phantom{-}1.55^{\text{****}}	&	\phantom{-}0.97^{\text{****}}	\Tstrut\\
\hline													
c_i	&	\phantom{-}1494^{\text{ }}	&	-2634^{\text{****}}	&	-11720^{\text{****}}	&	-4462^{\text{****}}	&	-2.47^{\text{****}}	&	-3.74^{\text{****}}	\Tstrut\\
h_i	&	\phantom{-}17913^{\text{****}}	&	\phantom{-}3518^{\text{****}}	&	\phantom{-}20606^{\text{****}}	&	\phantom{-}8083^{\text{****}}	&	\phantom{-}7.07^{\text{****}}	&	-0.24^{\text{ }}	\\
a_i	&	\phantom{-}114823^{\text{****}}	&	\phantom{-}43952^{\text{****}}	&	-8784^{\text{****}}	&	-2706^{\text{****}}	&	-3.94^{\text{****}}	&	\phantom{-}4.14^{\text{****}}	\\
\hline													
c_i h_i	&	-16104^{\text{****}}	&	\phantom{-}16649^{\text{****}}	&	\phantom{-}22008^{\text{****}}	&	\phantom{-}7930^{\text{**}}	&	\phantom{-}5.37^{\text{*}}	&	\phantom{-}13.81^{\text{****}}	\Tstrut\\
c_i a_i	&	\phantom{-}25205^{\text{****}}	&	-3695^{\text{*}}	&	-20315^{\text{***}}	&	-9147^{\text{***}}	&	-6.01^{\text{*}}	&	\phantom{-}0.02^{\text{ }}	\\
h_i a_i	&	-441920^{\text{****}}	&	-154036^{\text{****}}	&	-15849^{\text{*}}	&	-6867^{\text{*}}	&	-5.09^{\text{*}}	&	-9.14^{\text{****}}	\\
\hline													
\end{array}
\]
\label{tab:mlr-ind}
\end{table*}

\begin{table*}[tbp]
\caption{Multiple linear regression of a node's \emph{community-level} outcome measures on its update policy parameters ($c_i$, $h_i$, $a_i$) and their interactions (bottom three rows). Models were built on the same data as Table 1. Statistically significant coefficients are indicated with asterisks (*: $p < 10^{-2}$; **: $p < 10^{-3}$; ***: $p < 10^{-4}$; etc.).}
\[
\begin{array}{c|l|l|l|l|l}
\hline
\text{Outcome} & \multicolumn{1}{c|}{\text{community}} & \multicolumn{1}{c|}{\text{average}}     & \multicolumn{1}{c|}{\text{community}}    & \multicolumn{1}{c|}{\text{opinion}}    & \multicolumn{1}{c}{\text{opinion}} \\
\text{measure} & \multicolumn{1}{c|}{\text{size}}      & \multicolumn{1}{c|}{\text{edge weight}} & \multicolumn{1}{c|}{\text{deviance}} & \multicolumn{1}{c|}{\text{range}} & \multicolumn{1}{c}{\text{std.\ dev.}}\\
\hline
\text{const.} & \phantom{-}337.9^{\text{****}} & \phantom{-}4.99^{\text{****}} & \phantom{-}9.9^{\text{****}} & \phantom{-}1.65^{\text{****}} & \phantom{-}1.15^{\text{****}}  \Tstrut\\
\hline            
c_i & \phantom{-}65.3^{\text{****}} & -11.57^{\text{****}} & -3.67^{\text{****}} & -1.34^{\text{****}} & -2.49^{\text{****}}  \Tstrut\\
h_i & \phantom{-}21.3^{\text{ }} & -0.83^{\text{ }} & -0.53^{\text{ }} & -0.07^{\text{ }} & \phantom{-}4.33^{\text{****}}  \\
a_i & -68.1^{\text{****}} & \phantom{-}12.88^{\text{****}} & \phantom{-}3.57^{\text{****}} & \phantom{-}1.45^{\text{****}} & -0.07^{\text{ }}  \\
\hline            
c_i h_i & -295.8^{\text{****}} & \phantom{-}43.55^{\text{****}} & \phantom{-}14.12^{\text{****}} & \phantom{-}4.91^{\text{****}} & \phantom{-}4.14^{\text{*}}  \Tstrut\\
c_i a_i & \phantom{-}38.6^{\text{ }} & -0.28^{\text{ }} & \phantom{-}0.74^{\text{ }} & \phantom{-}0.06^{\text{ }} & \phantom{-}7.03^{\text{****}}  \\
h_i a_i & \phantom{-}118.5^{\text{ }} & -29.05^{\text{****}} & -8.57^{\text{****}} & -3.34^{\text{****}} & -15.21^{\text{****}}  \\
\hline            
\end{array}
\]
\label{tab:mlr-com}
\end{table*}

\begin{table*}[tbp]
\caption{Multiple linear regression of a node's \emph{community's assortativity} measures on its update policy parameters ($c_i$, $h_i$, $a_i$) and their interactions (bottom three rows). Models were built on the same data as Tables 1 and 2. Statistically significant coefficients are indicated with asterisks (*: $p < 10^{-2}$; **: $p < 10^{-3}$; ***: $p < 10^{-4}$; etc.).}
\[
\begin{array}{c|l|l|l|l}
\hline
\text{Outcome} & \multicolumn{1}{c|}{\text{assortativity}} & \multicolumn{1}{c|}{\text{assortativity on}} & \multicolumn{1}{c|}{\text{assortativity on}}    & \multicolumn{1}{c}{\text{assortativity on}} \\
\text{measure} & \multicolumn{1}{c|}{\text{on opinion}}      & \multicolumn{1}{c|}{\text{conformity ($c_i$)}} & \multicolumn{1}{c|}{\text{homophily ($h_i$)}} & \multicolumn{1}{c}{\text{neophily ($a_i$)}}\\
\hline
\text{const.} & \phantom{-}0.026^{\text{****}} & \phantom{-}0.004^{\text{****}} & \phantom{-}0^{\text{ }} & \phantom{-}0.026^{\text{****}}  \Tstrut\\
\hline          
c_i & -0.077^{\text{****}} & -0.014^{\text{****}} & -0.004^{\text{****}} & -0.047^{\text{****}}  \Tstrut\\
h_i & -0.006^{\text{ }} & -0.002^{\text{ }} & \phantom{-}0.001^{\text{ }} & -0.001^{\text{ }}  \\
a_i & \phantom{-}0.078^{\text{****}} & \phantom{-}0.013^{\text{****}} & \phantom{-}0.002^{\text{ }} & \phantom{-}0.054^{\text{****}}  \\
\hline          
c_i h_i & \phantom{-}0.269^{\text{****}} & \phantom{-}0.052^{\text{****}} & \phantom{-}0.004^{\text{ }} & \phantom{-}0.171^{\text{****}}  \Tstrut\\
c_i a_i & \phantom{-}0.021^{\text{ }} & \phantom{-}0.004^{\text{ }} & \phantom{-}0.013^{\text{*}} & \phantom{-}0.005^{\text{ }}  \\
h_i a_i & -0.163^{\text{****}} & -0.032^{\text{****}} & -0.003^{\text{ }} & -0.127^{\text{****}}  \\
\hline          
\end{array}
\]
\label{tab:mlr-pop}
\end{table*}

\begin{table*}[tbp]
\caption{Network level assortativity of weights $w_{ij}$ for various properties of upstream node $j$ and downstream node $i$. Calculations were based on the same data as Tables 1, 2 and 3. Assortativity coefficients significantly different from zero (one sample t-test, n=10) are indicated with asterisks (*: $p < 10^{-2}$; **: $p < 10^{-3}$; ***: $p < 10^{-4}$; etc.).}
\[
\begin{array}{c|l|l|l|l}
\hline
\text{Property} & \multicolumn{1}{c|}{x_j} & \multicolumn{1}{c|}{c_j} & \multicolumn{1}{c|}{h_j} & \multicolumn{1}{c}{a_j} \\
\hline
x_i & \phantom{-}0.0359^{\text{****}} &  \phantom{-}0.0004^{\text{ }}    &  \phantom{-}0.0011^{\text{ }}    &            -0.0011^{\text{ }}   \Tstrut\\   
c_i & \phantom{-}0.0005^{\text{ }}    &  \phantom{-}0.0077^{\text{****}} &  \phantom{-}0.0148^{\text{****}} &            -0.0226^{\text{****}}\\   
h_i & \phantom{-}0.0004^{\text{ }}    &            -0.0007^{\text{ }}    &            -0.0023^{\text{**}}   &  \phantom{-}0.0027^{\text{**}}  \\   
a_i & \phantom{-}0.0004^{\text{ }}    &           -0.0164^{\text{****}}  &            -0.0267^{\text{****}} &  \phantom{-}0.0407^{\text{****}}\\   
\hline          
\end{array}
\]
\label{tab:ass-uni}
\end{table*}

By comparison with low-conformity nodes, nodes with higher conformity tend to have significantly lower out-strength and community in-strength, have a less deviant, more conformist opinion relative to the entire network and relative to their own community, and belong to larger, less tight-knit, less deviant communities that tend to support a narrower range of opinion and be less assorted with respect to opinion, conformity, homophily or neophily. In short, conformity is associated with large, loose-knit, conformist communities with reduced opinion diversity.  

By comparison with low-homophily nodes, nodes with higher homophily tend to have significantly higher in-strength and out-strength, have a more deviant opinion relative to the entire network but not relative to their own community, and belong to communities that are more deviant relative to the entire network, but that are not distinctive in other respects, i.e., they are not significantly larger, smaller, more or less tight-knit, more or less diverse in terms of the range of opinion that they support, or more or less assorted by either opinion, conformity, homophily, or neophily. In short, homophily is associated with deviant, fragmented communities.  

By comparison with low-neophily nodes, nodes with higher neophily tend to have significantly higher in-strength but lower out-strength, have an opinion that is more conformist relative to the entire network but more deviant relative to their own community, and belong to communities that are neither more or less deviant, but are smaller, more tight-knit, more diverse in terms of the range of opinion that they support, and more assorted with respect to opinion, homophily and neophily. In short, neophily is associated with small, tight-knit, assorted communities with high opinion diversity; implying that highly neophilic nodes will tend to act as ``bridges'' between groups of nodes with divergent opinions and thereby serve to integrate the overall network to some degree.

Finally, in terms of network-wide assortativity (see Table 4), despite being significantly less strongly assorted on opinion than comparable Homogeneous networks, pairs of nodes in Uniform Heterogeneous networks are still significantly positively assorted with respect to their opinion ($x_j \asspos x_i$) due to the combined canalising effects of conformity and homophily tending to cause nodes with similar opinion to cluster together and nodes that cluster together to form similar opinions. This also results in significant positive assortativity with respect to conformity ($c_j \asspos c_i$). Nodes are also positively assorted with respect to neophily ($a_j \asspos a_i$) indicating that novelty seekers can tend to flock together. However, populations do not tend to positively assort on all elements of their update policy. Nodes are significantly negatively assorted with respect to their homophily ($h_j \assneg h_i$), since high homophily nodes will tend to separate from each other unless they happen to share a similar opinion. Node conformity and neophily are negatively assorted with each other in both directions ($c_j \assneg a_i$) reflecting the antagonism between these two tendencies. By contrast, node homophily is positively assorted with respect to node conformity for incoming edges only ($h_j \assposforward c_i$), i.e., high-homophily nodes tend to have more influence on highly conformist nodes. Finally, node homophily is negatively assorted with respect to node neophily for incoming edges but positively assorted for outgoing edges ($h_j \assnegforward a_i$, $a_j \assposforward h_i$), i.e., low-homophily nodes tend, predictably, to have more influence on highly neophilic novelty-seeking nodes, but, perhaps surprisingly, highly neophilic nodes tend to have more influence on highly homophilic nodes.

\subsection{Non-Uniform Heterogeneity}

So far, the three parameter values that determine a node's update policy have been drawn from the same uniform distribution, but this need not be the case. Here we explore the way in which altering the distribution of $c_i$, $h_i$ and $a_i$ values in the population causes network behaviour to depart from that described for the Uniform Heterogeneity case above.

We consider  a case designed to encourage networks to fragment into a larger number of smaller communities by biasing $h_i$ values to be relatively high compared to $c_i$ and $a_i$ values. We do this by drawing each of the three parameter values from normal distributions each with the same standard deviation, $\sigma_c=\sigma_h=\sigma_a=0.025$, but with parameter-specific means, $\mu_c=\mu_a=0.05$ and $\mu_h=0.25$. All parameter values are clipped to remain within the original range $[0.01, 0.3]$. That is, while each node's update policy will be allocated independently at random, a node's value of $h_i$ will tend to be at the higher end of the legal parameter range while its $c_i$ and $a_i$ values will tend to be at the lower end of this range. 

Under these conditions, and by comparison with the results reported in Tables 1--4, the influence of update policy on network formation is extremely attenuated. Equivalent multiple linear regression models to those presented in Tables 1--3 reveal that while a node's $a_i$ and, to a lesser extent, $h_i$ parameter values remain significant positive influences on its in-strength, and there remains a significant negative interaction between these two factors ($p < 10^{-5}$, $p < 10^{-3}$, and $p < 10^{-5}$, respectively), there are \emph{no} other significant main effects or interactions on \emph{any} other network property at individual, community or network level save that network nodes are now almost perfectly assorted on opinion ($r=0.999$, $p<10^{-36}$) and positively assorted on strength ($r=0.34$ by comparison with $r=-0.001$ in the Uniform Heterogeneous case). 

These results stem from the extremely strong fragmentary effect that high homophily imposes on the network. Despite nodes varying in all three of their update policy parameters, the dominating strength of homophily causes all nodes to rapidly form small isolated communities based on close opinion agreement; the average number of communities for these 10 networks is 50 by comparison with three in the Uniform Heterogeneous case. Differences between the relatively weak conformity and neophily tendencies of nodes have no significant influence on this process and hence these small communities are each effectively random with respect to anything other than node opinion. Node-level differences in the degree of neophily and homophily lead to small but significant differences in how strongly nodes are connected within each of their small communities, but the strength of a node's tendency towards social conformity has little influence as within-community opinions are effectively unanimous.

\section{Discussion}

The key findings reported here are first that uniform heterogeneity increases the extent to which extremism emerges in a population of adaptive social agents, and second that the way in which an agent's update policy differs from the norm has predictive power with respect to the properties of the network and community that they are eventually part of. This is the case despite the fact that initial conditions are uniform with respect to opinion and network topology, and that the model equations governing how agent opinions and connection weights change contain only linear terms. While the \emph{average} update policy in the Uniform Heterogeneous case and the comparable Homogeneous case are identical and the variation around this average is symmetrical for the Uniform Heterogeneous case (and nonexistent for the Homogeneous case), by contrast with what might be expected from a naive mean-field analysis of a well-mixed system, the ramifying effects of the self-structuring processes that take place on a network ensures that the outcomes of these two cases are divergent in significant respects. 

Perhaps encouragingly, although Heterogeneous conditions led to the presence of more extreme opinions and communities in the network, they did not lead to network fragmentation. Instead, populations formed under Uniform Heterogeneous conditions were more strongly integrated than either those formed under comparable Homogeneous conditions or those arising in (homogeneous) scenarios that exhibited strong extremism in previous studies \citep{sayama2020alife,sayama2022ppam}. This implies that although populations composed of agents with diverse update policies might be susceptible to extremist tendencies, it may be less challenging to ameliorate, resist or reverse the rise of extremism as even relatively extreme communities are likely to remain reachable by less extreme members of the population.

The current study leaves several issues unattended. First, several significant interaction effects are present in the multiple linear regression models reported here and their influence on network and opinion dynamics remains to be explained. Second, how might heterogeneity in the remaining model variables influence network and opinion dynamics? Agent-level heterogeneity in the two thresholds that govern homophily and neophily, $\theta_h$ and $\theta_a$, could be explored in order to address this question. 

Currently, the model assumes that an agent judges its similarly with other agents based only on the extent to which their expressed opinions are similar. Given that the extended model allows for agents to also differ in terms of the parameters governing social processes, there is the opportunity for similarity judgements to take into account the extent to which agents differ in terms of the update policies. However, it could be argued that an agent's update policy is not as directly observable as its opinion.   

Moreover, rather than initialising each agent with their own update policy and holding it fixed throughout network evolution, it would be interesting to consider endogenising update policy dynamics by modelling the way in which an agent's update policy parameters themselves might be influenced by exposure to interactions with neighbouring agents; something that is beginning to interest animal biologists under the topic of ``personality development'' \citep{SIH201550}. 

Finally, to date models of opinion dynamics on adaptive social networks have tended to represent the attitudes and opinions of each agent using a single scalar value (here referred to as the agent's opinion, denoted $x_i$). In reality, an agent's state is more complex; their interactions may influence and be influenced by attitudes and opinions that they hold on a number of different topics which may interact with each other in specific ways. Extending the present model to characterise agent state as a multi-dimensional vector would allow consideration of scenarios in which agents can be similar on some dimensions but dissimilar on others. For instance, how might the attitude and relationships between agents change based on the extent to which they share both social attitudes (e.g., a favourite football team) and political attitudes (e.g., a leaning towards left-wing or right-wing ideology). 

\section{Conclusions}

In this paper, we have extended an existing computational agent-based model of adaptive social network dynamics to feature agents that are heterogeneous in their update policies as well as their opinions and investigated the impact of this heterogeneity on the way in which extreme communities can arise through a combination of social conformity, homophily and neophily. Our results demonstrate that heterogeneity of this kind can have a systematic influence on the outcome of social network formation, but that the nature of this influence is sensitive to the structure of the heterogeneity involved. Relative to comparable Homogeneous conditions, networks formed under conditions of Uniform Heterogeneity tend to be more structurally integrated and less assorted with respect to agent opinion, but also tend to exhibit greater levels of extremism at both the individual and community level.

\section*{Acknowledgments}

HS was supported in part by JSPS KAKENHI Grant Number 19H04220 and 19K21571.



\footnotesize
\bibliographystyle{apalike}
\bibliography{bullock-sayama-alife2023}

\begin{thebibliography}{}

\bibitem[Badawy and Ferrara, 2018]{badawy2018rise}
Badawy, A. and Ferrara, E. (2018).
\newblock The rise of jihadist propaganda on social networks.
\newblock {\em Journal of Computational Social Science}, 1(2):453--470.

\bibitem[Bennett et~al., 2022]{bennett-alife2022}
Bennett, C., Lawry, J., and Bullock, S. (2022).
\newblock Exploiting intrinsic multi-agent heterogeneity for spatial
  interference reduction in an idealised foraging task.
\newblock In Holler, S., L\"offler, R., and Bartlett, S., editors, {\em
  Artificial Life 2022: Proceedings of the 2022 Artificial Life Conference}.
  MIT Press.

\bibitem[Blondel et~al., 2008]{blondel2008fast}
Blondel, V.~D., Guillaume, J.-L., Lambiotte, R., and Lefebvre, E. (2008).
\newblock Fast unfolding of communities in large networks.
\newblock {\em Journal of Statistical Mechanics: Theory and Experiment},
  2008(10):P10008.

\bibitem[B{\"o}hme and Gross, 2011]{bohme2011analytical}
B{\"o}hme, G.~A. and Gross, T. (2011).
\newblock Analytical calculation of fragmentation transitions in adaptive
  networks.
\newblock {\em Physical Review E}, 83(3):035101.

\bibitem[Bryden et~al., 2011]{eps271552}
Bryden, J., Funk, S., Geard, N., Bullock, S., and Jansen, V. (2011).
\newblock Stability in flux: Community structure in dynamic networks.
\newblock {\em Journal of The Royal Society Interface}, 8(60):1031--1040.

\bibitem[Cheng and Yu, 2019]{cheng2019}
Cheng, C. and Yu, C. (2019).
\newblock Opinion dynamics with bounded confidence and group pressure.
\newblock {\em Physica A}, 532:121900.

\bibitem[Coccia, 2016]{coccia2016radical}
Coccia, M. (2016).
\newblock Radical innovations as drivers of breakthroughs: characteristics and
  properties of the management of technology leading to superior organisational
  performance in the discovery process of {R\&D} labs.
\newblock {\em Technology Analysis \& Strategic Management}, 28(4):381--395.

\bibitem[Cole and Morgan, 2011]{cole2011vegaphobia}
Cole, M. and Morgan, K. (2011).
\newblock Vegaphobia: derogatory discourses of veganism and the reproduction of
  speciesism in {UK} national newspapers.
\newblock {\em The British Journal of Sociology}, 62(1):134--153.

\bibitem[Dall et~al., 2004]{dall2004}
Dall, S. R.~X., Houston, A.~I., and Mc{N}amara, J.~M. (2004).
\newblock The behavioural ecology of personality: consistent individual
  differences from an adaptive perspective.
\newblock {\em Ecology Letters}, 7:734--739.

\bibitem[Geard and Bullock, 2008]{eps266036}
Geard, N. and Bullock, S. (2008).
\newblock Group formation and social evolution: A computational model.
\newblock In Bullock, S., Noble, J., Watson, R., and Bedau, M., editors, {\em
  Artificial Life XI: Proceedings of the Eleventh International Conference on
  the Simulation and Synthesis of Living Systems}, pages 197--203. MIT Press.

\bibitem[Gross and Sayama, 2009]{gross2009adaptive}
Gross, T. and Sayama, H. (2009).
\newblock {\em Adaptive Networks}.
\newblock Springer.

\bibitem[Hegselmann and Krause, 2002]{hegselmann&krause2002}
Hegselmann, R. and Krause, U. (2002).
\newblock Opinion dynamics and bounded confidence: models, analysis and
  simulation.
\newblock {\em Journal of Artificial Societies and Social Simulation}, 5(3).

\bibitem[Holme and Newman, 2006]{holme2006nonequilibrium}
Holme, P. and Newman, M.~E. (2006).
\newblock Nonequilibrium phase transition in the coevolution of networks and
  opinions.
\newblock {\em Physical Review E}, 74(5):056108.

\bibitem[Jacyno et~al., 2013]{eps342217}
Jacyno, M., Bullock, S., Geard, N., Payne, T.~R., and Luck, M. (2013).
\newblock Self-organising agent communities for autonomic resource management.
\newblock {\em Adaptive Behavior}, 21(1):3--28.

\bibitem[Jacyno et~al., 2009]{eps267064}
Jacyno, M., Bullock, S., Luck, M., and Payne, T.~R. (2009).
\newblock Emergent service provisioning and demand estimation through
  self-organizing agent communities.
\newblock In Sierra, C., Castelfranchi, C., Decker, K.~S., and Sichman, J.~S.,
  editors, {\em Proceedings of the Eighth International Conference on
  Autonomous Agents and Multiagent Systems (AAMAS 2009)}, pages 481--488. ACM.

\bibitem[Kan et~al., 2023]{kan2023}
Kan, U., Feng, M., and Porter, M.~A. (2023).
\newblock An adaptive bounded confidence model of opinion dynamics on networks.
\newblock {\em Journal of Complex Networks}, 11(1):415–444.

\bibitem[Kata, 2012]{kata2012anti}
Kata, A. (2012).
\newblock Anti-vaccine activists, {Web} 2.0, and the postmodern paradigm--an
  overview of tactics and tropes used online by the anti-vaccination movement.
\newblock {\em Vaccine}, 30(25):3778--3789.

\bibitem[Kozma and Barrat, 2008]{kozma2008consensus}
Kozma, B. and Barrat, A. (2008).
\newblock Consensus formation on adaptive networks.
\newblock {\em Physical Review E}, 77(1):016102.

\bibitem[Levin et~al., 2021]{levin2021}
Levin, S.~A., Milner, H.~V., and Perrings, C. (2021).
\newblock The dynamics of political polarization.
\newblock {\em Proceedings of the National Academy of Sciences},
  118(50):e2116950118.

\bibitem[Liang et~al., 2013]{liang2013}
Liang, H., Yang, Y., and Wang, X. (2013).
\newblock Opinion dynamics in networks with heterogeneous confidence and
  influence.
\newblock {\em Physica A}, 392(9):2248--2256.

\bibitem[Lorenz, 2007]{lorenz2007}
Lorenz, J. (2007).
\newblock Continuous opinion dynamics under bounded confidence: A survey.
\newblock {\em International Journal of Modern Physics C}, 18(12):1819–1838.

\bibitem[Manrique et~al., 2018]{manrique2018generalized}
Manrique, P.~D., Zheng, M., Cao, Z., Restrepo, E.~M., and Johnson, N.~F.
  (2018).
\newblock Generalized gelation theory describes onset of online extremist
  support.
\newblock {\em Physical Review Letters}, 121(4):048301.

\bibitem[Morales et~al., 2015]{morales2015measuring}
Morales, A., Borondo, J., Losada, J.~C., and Benito, R.~M. (2015).
\newblock Measuring political polarization: Twitter shows the two sides of
  {Venezuela}.
\newblock {\em Chaos}, 25(3):033114.

\bibitem[Naranjo-Valencia et~al., 2017]{naranjo2017organizational}
Naranjo-Valencia, J.~C., Jimenez-Jimenez, D., and Sanz-Valle, R. (2017).
\newblock Organizational culture and radical innovation: Does innovative
  behavior mediate this relationship?
\newblock {\em Creativity and Innovation Management}, 26(4):407--417.

\bibitem[Pitonakova et~al., 2016a]{swarmintel2016}
Pitonakova, L., Crowder, R., and Bullock, S. (2016a).
\newblock Information flow principles for plasticity in foraging robot swarms.
\newblock {\em Swarm Intelligence}, 10(1):33--63.

\bibitem[Pitonakova et~al., 2016b]{alifexv-lenka}
Pitonakova, L., Crowder, R., and Bullock, S. (2016b).
\newblock Task allocation in foraging robot swarms: The role of information
  sharing.
\newblock In Gershenson, C., Froese, T., Siqueiros, J.~M., Aguilar, W.,
  Izquierdo, E.~J., and Sayama, H., editors, {\em Artificial Life XV:
  Proceedings of The Fifteenth International Conference on the Synthesis and
  Simulation of Living Systems}, pages 306--313. MIT Press.

\bibitem[Pitonakova et~al., 2018]{swarmintel2018}
Pitonakova, L., Crowder, R., and Bullock, S. (2018).
\newblock The {I}nformation-{C}ost-{R}eward framework for understanding robot
  swarm foraging.
\newblock {\em Swarm Intelligence}, 12(1):71--96.

\bibitem[Prior, 2013]{prior2013media}
Prior, M. (2013).
\newblock Media and political polarization.
\newblock {\em Annual Review of Political Science}, 16:101--127.

\bibitem[Reilly, 2016]{reilly2016gluten}
Reilly, N.~R. (2016).
\newblock The gluten-free diet: recognizing fact, fiction, and fad.
\newblock {\em The Journal of Pediatrics}, 175:206--210.

\bibitem[Sayama, 2009]{swarmchem}
Sayama, H. (2009).
\newblock Swarm chemistry.
\newblock {\em Artificial Life}, 15(1):105--114.

\bibitem[Sayama, 2020a]{sayama2020enhanced}
Sayama, H. (2020a).
\newblock Enhanced ability of information gathering may intensify disagreement
  among groups.
\newblock {\em Physical Review E}, 102:012303.

\bibitem[Sayama, 2020b]{sayama2020alife}
Sayama, H. (2020b).
\newblock Extreme ideas emerging from social conformity and homophily: An
  adaptive social network model.
\newblock In Bongard, J., Lovato, J., Hebert-Dufrésne, L., Dasari, R., and
  Soros, L., editors, {\em Proceedings of the 2020 International Conference on
  Artificial Life}, pages 113--120. MIT Press.

\bibitem[Sayama, 2022]{sayama2022ppam}
Sayama, H. (2022).
\newblock Social fragmentation transitions in large-scale parameter sweep
  simulations of adaptive social networks.
\newblock In {\em Proc.\ 14th International Conference on Parallel Processing
  and Applied Mathematics (PPAM 2022)}. Springer.

\bibitem[Sayama et~al., 2013]{sayama2013modeling}
Sayama, H., Pestov, I., Schmidt, J., Bush, B.~J., Wong, C., Yamanoi, J., and
  Gross, T. (2013).
\newblock Modeling complex systems with adaptive networks.
\newblock {\em Computers \& Mathematics with Applications}, 65(10):1645--1664.

\bibitem[Sayama and Sinatra, 2015]{sayama2015social}
Sayama, H. and Sinatra, R. (2015).
\newblock Social diffusion and global drift on networks.
\newblock {\em Physical Review E}, 91(3):032809.

\bibitem[Sayama and Yamanoi, 2020]{sayama2020beyond}
Sayama, H. and Yamanoi, J. (2020).
\newblock Beyond social fragmentation: Coexistence of cultural diversity and
  structural connectivity is possible with social constituent diversity.
\newblock In {\em Proceedings of NetSci-X 2020: Sixth International Winter
  School and Conference on Network Science}, pages 171--181. Springer.

\bibitem[Sih et~al., 2015]{SIH201550}
Sih, A., Mathot, K.~J., Moir\'on, M., Montiglio, P.-O., Wolf, M., and
  Dingemanse, N.~J. (2015).
\newblock Animal personality and state–behaviour feedbacks: a review and
  guide for empiricists.
\newblock {\em Trends in Ecology \& Evolution}, 30(1):50--60.

\bibitem[Yuan et~al., 2021]{yuan2021}
Yuan, Y., Yan, J., and Zhang, P. (2021).
\newblock {Assortativity measures for weighted and directed networks}.
\newblock {\em Journal of Complex Networks}, 9(2):cnab017.

\bibitem[Zanette and Gil, 2006]{zanette2006opinion}
Zanette, D.~H. and Gil, S. (2006).
\newblock Opinion spreading and agent segregation on evolving networks.
\newblock {\em Physica D}, 224:156--165.

\end{thebibliography}

\end{document}